\documentstyle[12pt]{article}
\begin{document}
\begin{titlepage}
\normalsize
\begin{center}
{\Large \bf Budker Institute of Nuclear Physics} 
\end{center}
\begin{flushright}
BINP 96-16\\

March 1996
\end{flushright}
\vspace{0.5cm}
\begin{center}
{\bf IS THE MODEL OF SPONTANEOUS $CP$-VIOLATION}
\end{center}
\begin{center}
{\bf IN THE HIGGS SECTOR}
\end{center}
\begin{center}
{\bf CONSISTENT WITH EXPERIMENT?}
\end{center}

\vspace{0.5cm}
\begin{center}
{\bf I.B. Khriplovich}\footnote{e-mail address: khriplovich@inp.nsk.su}
\end{center}
\begin{center}
Budker Institute of Nuclear Physics, 630090 Novosibirsk, Russia
\end{center}

\vspace{2.0cm}

\begin{abstract}
At natural values of parameters of the model dicussed, the
contribution of the chromoelectric dipole moment of the $s$-quark to
the neutron electric dipole moment (EDM) exceeds considerably the
experimental upper limit for the neutron EDM. As strict bounds on the
parameters of the model are derived from the atomic experiment with
$^{199}$Hg.
\end{abstract}

\end{titlepage}

{\bf 1.} The possibility of $CP$-violation being generated by the
spontaneous symmetry breaking in the Higgs fields interaction was
pointed out in \cite{lee}. A more realistic model based on this idea
was suggested later \cite{wei} and contains at least three doublets
of complex Higgs fields.

In the most ambitious approach one may try to ascribe to this
mechanism the $CP$-odd effects observed in $K$-meson decays. In this
case, however, not only the masses of charged Higgs bosons would be
rather low \cite{ad,au}. Various estimates for the neutron EDM in
this version \cite{zk,kz,abgu,kkz} lead to the predictions: 
\begin{equation}
d(n)/e \sim 10^{-24} - 10^{-23}\,\mbox{cm},
\end{equation}
well above the experimental upper limit \cite{sm,al}:
\begin{equation}\label{ex}
d(n)/e\,<\,7\cdot 10^{-26}\,\mbox{cm},
\end{equation}

But then one can pass over to a more "natural" version of this model, 
with heavy Higgs bosons. Of course, in this case the model is
responsible for only a small portion of $CP$-violation in kaon decays. 
It would be new physics, a new source of $CP$-violation, supplemental
to that generating the effects already observed. 

The dominant contribution to the dipole moments in this model is given
by diagrams of the type 1 with a heavy particle ($t$-quark, $W$-boson
or Higgs) propagating in the upper loop \cite{bz}.  For the neutron
dipole moment this approach is further elaborated upon in
\cite{gw,cky,hmp}. In particular, it is pointed out there that, in
the model discussed, the neutron EDM is controlled by diagram 2 with
the $t$-quark propagating in the upper loop, but both wavy lines
corresponding to gluons. The effective operator generated by this
diagram is
\begin{equation}\label{c}
H_{c}=\,\frac{1}{2}\,d^c\,\bar q \gamma_5\sigma_{\mu\nu}t^a q\,G^a_{\mu\nu}
\end{equation}
where $t^a\,=\,\lambda^a/2$ are the generators of the colour $SU(3)$
group. The constant $d^c$ in expression (\ref{c}) is called the
quark chromoelectric dipole moment (CEDM).

The value of the $d$-quark CEDM, as obtained directly from diagram 1, 
is \cite{gw,cky}
\begin{equation}\label{c1}
d^c\,=\,g_s\,\frac{G}{\sqrt 2}\,m_d\,\frac{\alpha_s}{16 \pi^3}\,
\{\,\mbox{Im} Z_0[f(z)+g(z)]\,-\,\mbox{Im} \tilde{Z}_0[f(z)-g(z)]\}.
\end{equation}
In this expression $g_s$ is the quark-gluon coupling constant, 
$\alpha_s=\,g_s^2/4\pi,\,\, G$ is the Fermi weak interaction
constant, $m_d$ is the quark mass, Im$Z_0$ and Im$\tilde{Z}_0$ are
$CP$-violating parameters of the model. Functions $f$ and $g$
describe the CEDM dependence on the ratio of the $t$-quark mass to
the mass of the lightest neutral Higgs boson,
$z\,=\,m_t^2/M_{H^0}^2$. At $z\sim 1$ both functions are close to
unity. Their general $z$ dependence is given in Refs. \cite{bz,cky}.
An analogous expression was derived in Refs. \cite{gw,cky} for the 
$u$-quark CEDM.

To investigate the CEDM contribution to the observable effects, we
have to bring the expressions (\ref{c}), (\ref{c1}) down from the
scale of $M\,\sim\,200$ GeV to the usual hadronic scale $m\,\sim 1$
GeV. In particular, to substitute for $m_d$ the usual current mass
value $7$ MeV, we have to introduce the renormalization group (RG)
factor 
$$\left[\frac{\alpha_s(M)}{\alpha_s(m)}\right]^{12/23}.$$ 
Now, the QCD sum rule technique, used below to estimate the CEDM
contribution to observable effects, is applied directly to the
operators of the type 
$$g_s\bar q\,\gamma_5\sigma_{\mu\nu}t^a q\,G^a_{\mu\nu},$$ 
which include $g_s$ explicitly. This brings one more
RG factor \cite{svz}
$$\left[\frac{\alpha_s(M)}{\alpha_s(m)}\right]^{2/23}.$$ 
On the other hand, as distinct from Refs. \cite{gw,cky}, we see no
special reasons to bring the explicit $\alpha_s$ factor, entering the
expression (\ref{c1}), down from the high-momenta scale $M$, where it
is defined at least as well as at $m\,\sim\,1$ GeV.  The overall RG
factor, introduced in this way into formula (\ref{c1}), is
\begin{equation}
\left[\frac{\alpha_s(M)}{\alpha_s(m)}\right]^{14/23}
\end{equation}

Now, assuming
$$\mbox{Im}Z_0[f(z)+g(z)]\,-\,\mbox{Im}\tilde{Z}_0[f(z)-g(z)]\,\sim\,1,$$
we arrive at the following numerical estimate for the quark CEDM:
\begin{equation}\label{ces}
d^c\, \sim\, 3\cdot 10^{-25}\,\mbox{cm}.
\end{equation}

\bigskip
 
{\bf 2.} However, the most serious problem is to find the CEDM contribution  
to the neutron dipole moment. Here our conclusions differ from
those of Refs. \cite{gw,cky}. The simplest way \cite{kk} to estimate this
contribution is to assume, just by dimensional reasons, that $d(n)/e$ is
roughly equal to $d^c$ (obviously, the electric charge $e$ should be
singled out of $d(n)$, being a parameter unrelated to the nucleon structure). 

In a more elaborate approach \cite{kk}, the neutron EDM is estimated in the 
chiral limit via diagram 3, according to Ref. \cite{cdv}. For both $u$- and
$d$-quarks, the contribution of operator (\ref{c}) to the $CP$-odd $\pi
NN$ constant $\bar{g}_{\pi NN}$ is transformed by the PCAC technique to the same
expression:
\begin{equation}\label{pi}
<\,\pi^- p\,|\,g_s\bar q \gamma_5\sigma_{\mu\nu}t^a
q\,G^a_{\mu\nu}|\,n\,> = \, \frac{i}{f_{\pi}}\,
 <\,p\,|\,g_s\bar u \sigma_{\mu\nu}t^a d\,G^a_{\mu\nu}\,|\,n\,>. 
\end{equation}
We include the quark-gluon coupling constant $g_s$ explicitly into the
above relation since the corresponding estimate based on the QCD sum
rules refers directly to the last matrix element. This estimate gives
a value close to\newline 
$-1.5$ GeV$^2$. For momenta $\sim 1$ GeV in this
estimate, we take $g\approx 2$. Then the result for the neutron EDM
is:
\begin{equation}
d(n)/e\,\sim\,2\cdot 10^{-25}\,\mbox{cm},
\end{equation}
which exceeds the experimental upper limit (\ref{ex}).

Let us introduce the ratio of the neutron dipole moment, as induced by a CEDM,
to $d^c$ itself:
\begin{equation}\label{ro}
\rho\,=\,\frac{d(n)/e}{d^c(q)}.
\end{equation}
Its value obtained in this approach, $\rho = 0.7$, is quite close indeed to
unity. In our opinion, this good agreement with the above simple-minded
result enhances the reliability of both estimates.

A quite essential contribution to the neutron EDM can be induced by the
chromoelectric dipole moment $d^c(s)$ of the $s$-quark \cite{hmp}. The
gain in the magnitude of $d^c(s)$, as compared to the $d$-quark CEDM,
is the large ratio of the quark masses, $m_s/m_d\approx 20$.

On the other hand, for the $s$-quark, the ratio 
\begin{equation}\label{ros}
\rho_s\,=\,\frac{d(n)/e}{d^c(s)}.
\end{equation}
should be much smaller than unity. Indeed, according to the QCD sum
rule calculations of Ref. \cite{kkz}, it is about 0.1. One should
mention that other estimates \cite{zk,hmp1} predict for the ratio
(\ref{ros}) a value an order of magnitude smaller, and this smaller
prediction was used in Ref. \cite{hmp}.

Then, how reliable is the estimate $\rho_s\,=\,0.1$? There are strong
indications now that the admixture of the $\bar s s$ pairs in
nucleons is quite considerable.  In particular, it refers to the spin
content of a nucleon. And though these indications refer to operators
different from $\bar s \gamma_5\sigma_{\mu\nu}\,t^a s\,G^a_{\mu\nu}$,
they give serious reasons to believe that the estimate
\begin{equation}
\rho_s\,=\,0.1
\end{equation}
is just a conservative one.

At this value of $\rho_s$ the resulting contribution of the $s$-quark
CEDM to the neutron dipole moment
\begin{equation}\label{}
d(n)/e\,=\,6 \cdot 10^{-25}\,\mbox{cm}
\end{equation} 
is larger than the experimental upper limit (\ref{ex}) almost by an
order of magnitude.

\bigskip

{\bf 3.} At last, let us compare the predictions of the model
discussed with the result of the atomic experiment. The measurements of
the EDM of the mercury isotope $^{199}$Hg have resulted \cite{lam} in
\begin{equation}\label{hg}
d(^{199}\mbox{Hg})/e\,<\,9\cdot 10^{-28}\mbox{cm}.
\end{equation}
According to calculations of Ref. \cite{kky}, it corresponds to the
upper limit on the $d$-quark CEDM 
\begin{equation}\label{}
d^c\,<\,2.4\cdot 10^{-26}\mbox{cm}
\end{equation}
The prediction (\ref{ces}) exceeds this upper limit by an order of
magnitude.

Our analysis demonstrates that very special assumptions concerning
the parameters of the model of spontaneous $CP$-violation in the
Higgs sector (such as large mass $M_{H^0}$ of the Higgs boson, small
values of the $CP$-violating parameters Im$Z_0$, Im$\tilde{Z}^0$,
etc) are necessary to reconcile the predictions of this model with
the experimental upper limits on the electric dipole moments of
neutron and $^{199}$Hg.

Such fine tuning will change as well the prediction of the model 
for the electron EDM. It will make much smaller the accepted now
prediction\newline $d(e)\,\sim\,10^{-27}$ cm \cite{bz,gv,lp}, which
is only an order of magnitude below the present experimental upper
limit \cite{com}.

\bigskip
\bigskip
\bigskip
\bigskip

I am grateful to J. Ellis, P. Herczeg and S.K. Lamoreaux for the
discussions of results. The investigation was supported by the
Russian Foundation for Basic Research through grant No.95-02-04436-a,
and by the National Science Foundation through a grant to the
Institute for Theoretical Atomic and Molecular Physics at Harvard
University and Smithsonian Astrophysical Observatory. 

\end{document}